# Light tuning of selective Bragg diffraction of the oblique helicoidal cholesteric

IGOR GVOZDOVSKYY,[1,*] HALYNA BOGATYRYOVA,[1] NATALIA KASIAN,[2,3] LONGIN LISETSKI,[3] AND VITALII CHORNOUS[4]

[1]*Optical Quantum Electronics Department, Institute of Physics of the National Academy of Sciences of Ukraine, 46 Nauky Avenue, Kyiv, 03028, Ukraine*

[2]*Physics Faculty, Warsaw University of Technology, 75 Koszykowa Street, Warsaw, 00-662, Poland*

[3]*Nanostructured Materials Department, Institute for Scintillation Materials of STC "Institute for Single Crystals" of the National Academy of Sciences of Ukraine, 60 Nauky Avenue, Kharkiv, 61072, Ukraine*

[4]*Medical and Pharmaceutical Chemistry Department, Bukovinian State Medical University, 2 Teatral'na Avenue, Chernivtsi, 58002, Ukraine*

*\*igvozd@gmail.com*

**Abstract:** It is known that the oblique helicoidal cholesteric structure ($Ch_{OH}$), which is formed in alternating electrical field applied to the chiral nematic phase ($N^*$) at temperatures above the chiral twist-bend nematic phase ($N^*_{tb}$), can show selective Bragg reflection in the visible range of spectrum. In this work, we study the effects of AC electric field at different frequencies, as well as UV irradiation on the Bragg reflection of $Ch_{OH}$ structure of the $N^*_{tb}$-forming mixture containing a light sensitive chiral azo-compound. It has been found that $Ch_{OH}$ structure shows two sequential states in the tuning of selective Bragg reflection at high and low electrical fields.

1. Introduction

Cholesteric liquid crystals (CLCs) are characterized by a periodic helicoidal structure in which the preferred local orientation of long axes of the molecules (so-called director $\bar{n}$) in quasi-nematic layers is perpendicular to the helical axis and rotates by a certain angle in transition between the adjacent quasi-nematic layers [1]. This periodic structure is formed either by chiral mesogenic compounds like cholesterol esters, *i.e.*, predominantly, by chiral molecules (so-called chiral dopant or for short ChD) added to the nematic host. CLCs are characterized by the cholesteric helix pitch $P_0$, which depends on both the helical twisting power $β$ (HTP) and concentration of ChD. The pitch of cholesteric helix can change within a wide range of values, from centimeters to nanometers [1,2]. Depending on the nature of interaction between the molecules of the nematic host and chiral dopant, the cholesteric helix can be right or left-handed, which leads to separate reflection and transmission of left or right-handed circular polarized light. If the helicoidal structure selectively reflects light around the maximum wavelength $λ_{max}$ in the visible spectral range (so-called selective Bragg reflection of light (BRL)), at normal incidence of light the pitch of cholesteric helix is

$P_0 = \lambda_{max}/\langle n \rangle$, where $\langle n \rangle = (n_o + n_e)/2$ is the average refractive index, determined by the ordinary $n_o$ and extraordinary $n_e$ refractive indices of the medium [1]. The sensitivity of the cholesteric helix pitch to different external influences (*e.g.* temperature [3,4], magnetic and electric fields [5-8], light [9-13], vapors and gases [14-16]), makes CLCs highly attractive for different optical applications, such as filters and mirrors [17-19], lasers [20-22], smart windows and display devices [23]. Unfortunately, the tuning of the Bragg reflection wavelength by means of the applied electric field for cholesteric helicoidal structure is significantly limited, as was explained in [24]. However, there exist successful applications for long helix pitch cholesterics in electric field when no tuning of BRL is observed. For example, at the certain ratio between the thickness of liquid crystal (LC) layer *d* and length of helix pitch $P_0$, the Helfrich-Hurault instabilities [6,7] are observed that can be efficiently used to obtain Raman-Nath diffraction gratings [25] with electrically tunable period [26-28].

In the last decade much attention is paid to studies of the theoretically predicted in [29-31] novel mesophase (so-called twist-bend nematic phase or $N_{tb}$), formed by non-chiral flexible dimer (or banana-shaped [31]) molecules, having oblique helicoidal helix with nanoscale pitch about 8 nm, and firstly experimentally described in [32-34]. Chiral twist-bend nematic ($N^*_{tb}$) phase obtained by adding of ChDs to $N_{tb}$ phase was firstly studied in an electric field [35,36]. A special cholesteric state (so-called *oblique helicoidal cholesteric* or $Ch_{OH}$) is formed by applying the electric field to the chiral phase ($N^*$) of the $N_{tb}$-forming mixture doped with small concentration of ChD, showing the phenomenon of selective BRL in a wide spectral range from ultraviolet (UV) to infrared (IR) [36]. Changes in the main characteristics of the applied electric field, such as the voltage applied to a liquid crystal cell of a certain thickness (*i.e.*, electric field strength) [36-38] and the AC field frequency [39,40] allow the tuning of the Bragg reflection wavelength, which makes these liquid crystal materials very promising in various applications [37,39,41].

In this manuscript, we propose additional means for tuning the wavelength of selective Bragg reflection on $Ch_{OH}$ at oblique incidence of light using UV radiation, which becomes possible owing to the use of chiral azo-compound ChD-3816 [42], as a constituent component of the $N^*_{tb}$-forming system.

## 2. Photosensitive $Ch_{OH}$ and experimental studies

*2.1 Materials and methods*

Figure 1a shows the structure of the molecule (1*R*,2*S*,5*R*)-2-isopropyl-5-methylcyclohexyl-4-{(*E*)-[4-(hexanoyloxy)phenyl]diazenyl}benzoate (or ChD-3816 for short) as a left-handed light-sensitive chiral dopant, the synthesis of which was described in detail elsewhere [42]. 10 wt.-% of ChD-3816 was dissolved in the basic $N_{tb}$-forming mixture (Figure 1b), consisting of two achiral twist-bend nematic dimers CB7CB (synthesized in Bukovinian State Medical University, Chernivtsi, Ukraine), CB6OCB (Synthon Chemicals GmbH & Co, Wolfen, Germany) and 5CB (STC "Institute of Single Crystals", Kharkiv, Ukraine) in the weight ratio (39:19:42). This mixture possesses a wide temperature range of the $N^*$ phase, where $Ch_{OH}$ structure with selective BRL in the visible light spectrum were observed in the electric field. The characterization of mixtures contain various concentrations of ChD-3816 was carried out in [42].

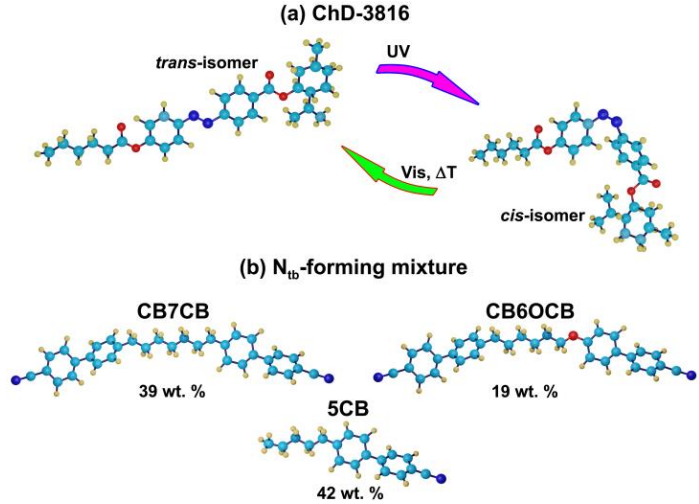

Fig. 1. (a) The molecular structure of (1*R*,2*S*,5*R*)-2-isopropyl-5-methylcyclohexyl-4-{(*E*)-[4-(hexanoyloxy)phenyl]diazenyl}benzoate (ChD-3816) in *trans* and *cis*-forms. (b) $N_{tb}$-forming mixture consisting of two achiral twist-bend nematic dimers CB7CB, CB6OCB and 5CB in the weight ratio 39:19:42, respectively.

The LC cell was assembled from two glass substrates coated with both conductive layer ITO (Indium Tin Oxide) and polyimide polymer PI2555 (HD MicroSystems, USA). The thin polyimide aligning layers were unidirectionally rubbed $N_{rubb}$ = 15 times to provide their strong azimuthal anchoring energy [43].

The thickness of LC cell, measured by the transmission spectrum of the empty cell, was about 25 μm. The LC cell was filled by the $N^*_{tb}$ phase that should show the $Ch_{OH}$ structure in the electric field.

Characteristics of the photosensitive $Ch_{OH}$ containing 10 wt. % ChD-3816 were described in [42,44].

*2.2 Two states of selective BRL at oblique incidence of light on $Ch_{OH}$ structure*

The typical behavior of $Ch_{OH}$ in an alternating electric field is the unwinding of pitch $P$ of the $Ch_{OH}$ helix with decreasing of voltage $U$, which is expressed as follows [29]:

$$P = \frac{2\pi \cdot d}{U} \sqrt{\frac{K_{33}}{\varepsilon_0 \cdot \Delta\varepsilon}} \qquad (1)$$

where $\varepsilon_0$ is the constant of vacuum permittivity and $\Delta\varepsilon$ is the dielectric anisotropy, $d$ is the thickness of LC cell, $K_{33}$ is the bend elastic constant.

In this case the decrease of applied voltage $U$ leads to increase of reflected wavelength $\lambda$, *i.e.* the so-called red shift is observed [35,36]. The electrical tuning of the wavelength of the maximum selective BRL in the visible range of spectrum is promising for different applications in optics [39].

To observe the selective BRL of $Ch_{OH}$ structure in an alternating electric field, the light from tungsten lamp is incident at a certain angle α to the normal of the LC cell (Figure 2). The recording of transmission spectra of the $Ch_{OH}$ in LC cell was carried out by an Ocean Optics USB400 (California, USA) spectrometer for a certain frequency $f$ of the applied field at various values of voltage $U$.

However, it should be noted that the tuning of the selective BRL at oblique incident on $Ch_{OH}$ with changing of voltage $U$ at certain fixed frequency $f$ of electric field was described in detail elsewhere [35-38,44].

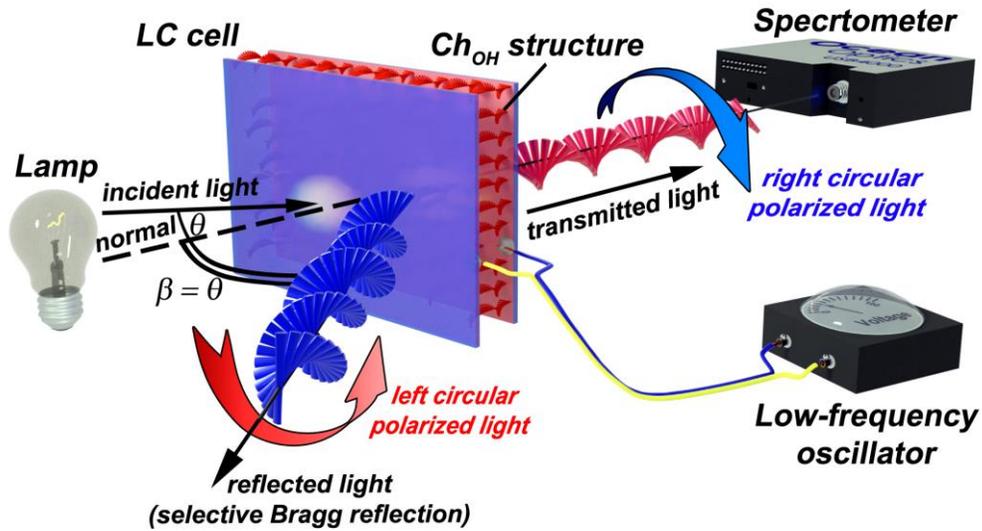

Fig. 2. Scheme of the observation of the selective Bragg reflection of light incident at certain angle $\theta$ to the normal of the LC cell and the recording of transmission spectrum of the LC cell filled by the $N^*_{tb}$-forming mixture characterized by $Ch_{OH}$ structure under alternating electrical field.

Thus, at very high voltage the homeotropic (vertical) orientation of the nematic phase (N) is observed. Decrease of voltage leads to the appearance of $Ch_{OH}$ having the BRL at short wavelengths of the visible spectrum range. The threshold voltage $U_{NC}$ of this transition is expressed as follows [35,36]:

$$U_{NC} = (2\pi \cdot d/P_0)(K_{22}/\sqrt{\varepsilon_0 \cdot \Delta\varepsilon \cdot K_{33}}) \qquad (2)$$

where $P_0$ is the length of cholesteric pitch of $N^*$ phase of the $N^*_{tb}$-forming mixture, $d$ is the thickness of LC cell, $\varepsilon_0$ is the constant of vacuum permittivity and $\Delta\varepsilon$ is the dielectric anisotropy, $K_{22}$ and $K_{33}$ is the twist and bend elastic constants, respectively.

Further decreasing of the voltage $U < U_{NC}$ leads to the red shift of selective BRL wavelength.

Figure 3a shows sequential changes of transmission spectrum (*i.e.* from spectrum 1 to spectrum 8) of the LC cell depending on applied voltage $U$. This dependence is the same as described in detail elsewhere [35,36]. Figures 3b, c and d show photographs of LC cell at decreasing voltage $U$. At certain applied voltage $U$, Ch$_{OH}$ possesses selective reflection of light with a certain frequency $\nu = c/\lambda$ which is responsible for the color of LC cell (where c is the speed of light in vacuum).

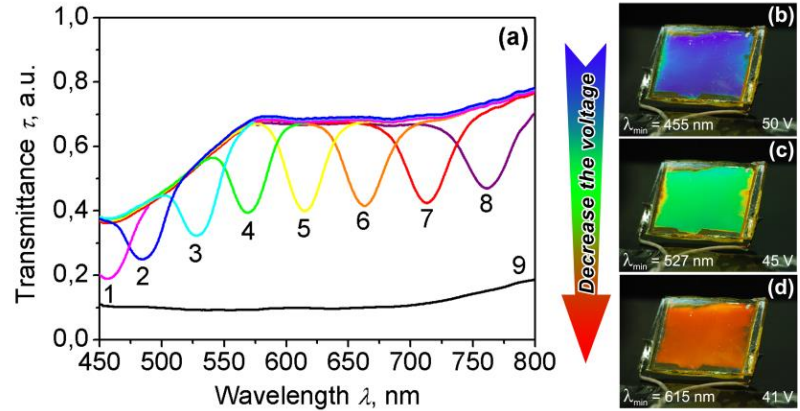

Fig. 3. (a) Sequential changes of the transmission spectrum in the visible light range of the Ch$_{OH}$ under decrease of the applied voltage $U$: (1) – $\lambda_{min}$ = 457 nm at 50.7 V; (2) – $\lambda_{min}$ = 485 nm at 47 V; (3) – $\lambda_{min}$ = 527 nm at 45 V; (4) – $\lambda_{min}$ = 570 nm at 43 V; (5) – $\lambda_{min}$ = 615 nm at 41 V; (6) - $\lambda_{min}$ = 660 nm at 39 V; (7) – $\lambda_{min}$ = 715 nm at 37 V; (8) – $\lambda_{min}$ = 760 nm at 35 V. Spectrum 9 corresponds to the N phase at 25 V after Ch$_{OH}$ - nematic transition. Photography of the LC cell under alternating of electrical field with frequency $f$ = 1 kHz and certain voltage $U$: (b) 50 V; (c) 43 V and (d) 41 V.

In a recent study [44] it has been shown that at oblique incidence of light on Ch$_{OH}$, at least two states of successive changes of the BRL are observed when the applied electric field is decreased, namely at high (State 1) and low (State 2) voltages $U$. These states were ascribed to different orders $m_i$ of Bragg diffraction in the visible range of spectrum. At oblique incidence of light on Ch$_{OH}$ layer with the decrease of voltage the sequential appearance of 1$^{st}$, 2$^{nd}$ and 3$^{rd}$ order ($m_i$) of the diffraction of Bragg is happening. Based on the Equation 1 and Equation 2 the relationship between maximum of wavelength $\lambda_{max}$ of the BRL and applied voltage $U$ is expressed as follows [44]:

$$\lambda = \frac{2 \cdot d \cdot \langle n \rangle \cdot \sin(\theta)}{U \cdot m_i} \frac{K_{22}}{\sqrt{\varepsilon_0 \cdot \Delta\varepsilon \cdot K_{33}}} \quad (3),$$

where $\theta$ is incident angle of the light on Ch$_{OH}$ structure.

In Ref. [44] it was also noted that the observation of 1st order of the BRL is complicated due to the need to apply very high voltage that may lead to electrical breakdown of the LC cell.

Figure 4 shows the experimental data on the sequential unwinding of helical pitch $P$ of the Ch$_{OH}$ leading to the red shift of Bragg reflection wavelength with the decrease of value of the applied field for both high (State 1, red solid circles) and low (State 2, blue solid squares) voltage. By taking into account the Equation (3), the calculated data for the 2nd and 3rd order of the Bragg diffraction are shown by the blue dashed curve 1 and red dash-dotted curve 2, respectively.

At a certain voltage $U$, the jump of the Bragg reflection wavelength occurs to its initial value observed at high voltage. In this case the jump of wavelength is accompanied by the blue shift of the Bragg reflection wavelength (Figure 4, State 2, blue solid squares). In addition, when the Ch$_{OH}$ is in so-called "*hybrid state*" ($U \sim 43$ V), two peaks appear in the spectrum of BRL, as it was found recently [44].

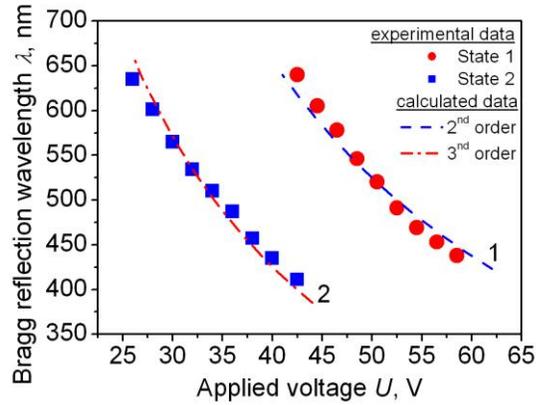

Fig. 4. Dependence of the selective Bragg reflection wavelength λ on high (red solid circles) and low (blue solid squares) voltage $U$ applied to the LC cell filled by N*$_{tb}$-forming mixture containing 10 wt. % ChD-3816. The calculated dependencies for: 1) 2nd order - blue dashed curve1 and 2) 3rd order - red dash-dotted curve 2.

If the applied field continues to decrease, then at a certain threshold voltage $U_{N*C} < U$ the transition of the Ch$_{OH}$ structure to N* phase occurs, which is accompanied by the reorientation of the oblique helicoidal axis to lying axis. This threshold electric field is expressed as: [29]

$$U_{N*C} \approx U_{NC} \cdot \frac{K_{33}}{K_{22} + K_{33}} [2 + \sqrt{2 \cdot (1 - \frac{K_{33}}{K_{22}})}] \quad (4)$$

The Ch$_{OH}$ – N* transition is accompanied by light scattering, and accordingly a decrease in the LC cell transmittance is observed (spectrum 9 in Figure 3a).

## 2.3 Effects of UV irradiation on Ch$_{OH}$ structure of the photosensitive N*$_{tb}$-forming mixture

It is obvious that not less intriguing and interesting is another feature that could be used with photosensitive N*$_{tb}$-forming mixtures, namely, the influence of UV light on spectral characteristics of Ch$_{OH}$ structure.

We have found that the action of UV irradiation on Ch$_{OH}$ structure leads to an increase in the voltage $U$ needed for tuning of wavelength of the BRL maximum for both the State 1 and State 2.

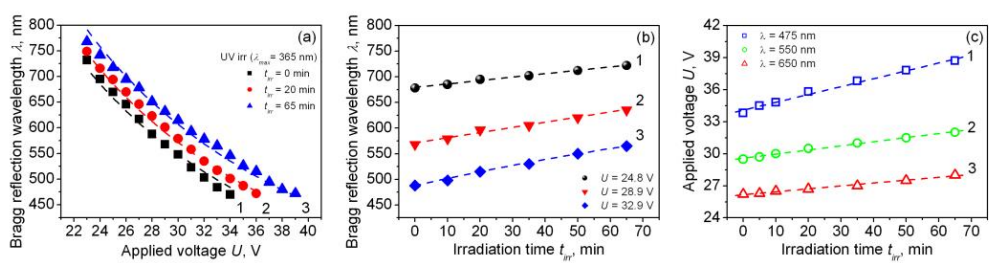

Fig. 5. Dependence of the selective BRL wavelength $\lambda_{max}$ on: (a) applied voltage $U$ and (b) UV irradiation time $t_{irr}$, for the State 2 of the LC cell filled by N*$_{tb}$-forming mixture containing of the 10 wt. % ChD-3816. (c) Irradiation time dependence of the value of the applied voltage $U$ for the specified values of the wavelength of BRL: (1) - 475 nm (opened blue squares), (2) - 550 nm (opened green circles) and (3) - 650 nm (opened red triangles). The thickness of LC cell is 25.2 µm. The frequency $f$ of the applied field is 1 kHz.

Figure 5a shows the dependence wavelength of selective BRL maximum on the applied voltage $U$ for different UV irradiation times $t_{irr}$ in case of the 3$^{rd}$ order of Bragg diffraction (*i.e.* State 2) that occurs using photosensitive N*$_{tb}$-forming mixture with chiral azo-compound ChD-3816 [42].

UV irradiation leads to the decrease of amount of the *trans*-isomer of ChD-3816, which possesses higher HTP than the *cis*-isomer [42]. The *trans-cis* photoisomerization causes the red shift of both the wavelength of the selective BRL maximum (Figure 5b) and voltage $U$ needed to preserve the constant value of the wavelength $\lambda_{max}$ of BRL (Figure 5c).

Figure 6 shows the LC cell with Ch$_{OH}$ structure for different values of applied voltage $U$ and the constant frequency $f$ irradiated by UV light through mask having a hole with diameter 5 mm.

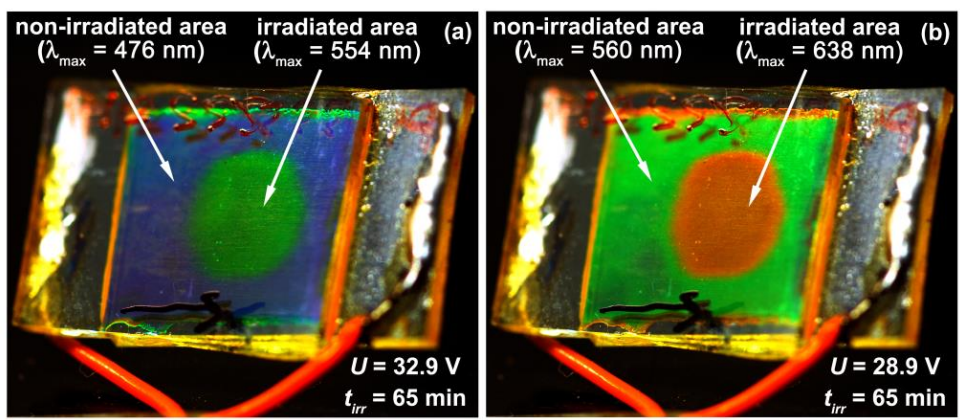

Fig. 6. Photographs of LC cell with Ch$_{OH}$ structure for different values of voltage $U$: a) 32.9 V and b) 28.9 V and applied field frequency $f = 1$ kHz, characterized by different wavelengths $\lambda_{max}$ of the selective BRL for non-irradiated (background) and UV irradiated (circle) areas.

For instance, Figure 7 shows the Ch$_{OH}$ structure which is in State 2, *i.e.*, when 3$^{rd}$ order of Bragg diffraction is observed. At the beginning of the study of the reversible *cis-trans* isomerization both under Vis irradiation and influence of temperature, the N$^*_{tb}$-forming mixture (curves 1, opened black circles, Figure 7) was irradiated by UV light with $\lambda_{max} = 365$ nm for $t_{irr} = 65$ min. After UV irradiation of Ch$_{OH}$ structure the shift of the applied voltage $U$ value (curves 2, opened blue triangles, Figure 7) is observed. This state of Ch$_{OH}$ structure is the initial condition to study the reversible photoisomerization induced by Vis irradiation or by the effects of temperature. Using the following parameters: $m_i = 3$, $d = 25.2$ μm, $K_{22} = 3 \times 10^{-12}$ pN, $\langle n \rangle = 1.7$ the curve 1 (black color) and curve 2 (blue color) were calculated according to Equation (3) for incidence angles $\theta = \pi/3.3$ and $\theta = \pi/2.8$, respectively. The good agreement between the measured data and curves calculated using Equation (3) is observed.

Owing to the irradiation by Vis (*e.g.* by means of the incandescent lamp) or thermal impact (*e.g.* storage at $T = 80$ °C) the reversible *cis-trans* isomerization of the ChD-3816 molecules (as reported in ethanol solution [42]) leads to the restoration of the N$^*_{tb}$-forming mixture in its original state, for instance in case of the Stage 2 (*i.e.* 3$^{rd}$ order of Bragg diffraction [44]), as shown in Figure 7a and Figure 7b, respectively.

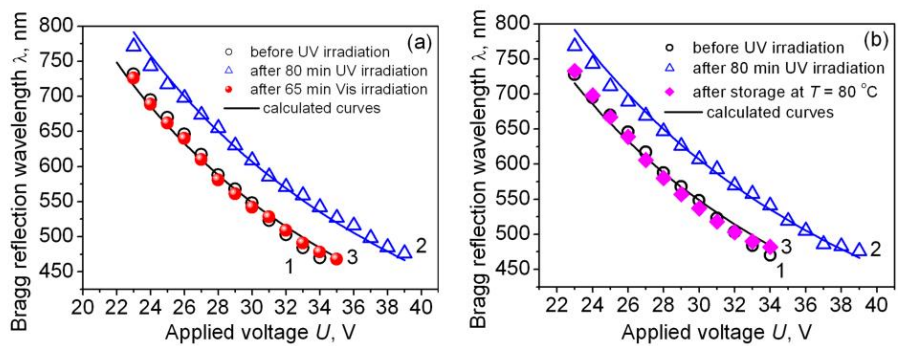

Fig. 7. Dependence of the Bragg reflection wavelength $\lambda$ on the voltage applied to LC cell, filled by $N^*_{tb}$-forming mixture (curves 1, opened black circles) and additionally irradiated by UV light (curves 2, opened blue triangles), in the case: (a) Vis irradiation for 65 min (curve 3, red spheres) and (b) storage at $T = 80$ °C within 60 min (curve 3, solid magenta diamonds). The frequency $f$ of the applied field is 1 kHz.

The *cis-trans* isomerization of ChD-3816 molecules caused by the Vis irradiation for $t_{irr} = 65$ min (red spheres, curve 3) or storage at $T = 80$ °C within $t_{stor} = 60$ min (solid magenta diamonds, curve 3) of the $N^*_{tb}$-forming mixture leads to restoration of the initial state (curves 1), as shown in Figure 7.

It is important to note that at the reversible *cis-trans* isomerization which is a consequence of both the Vis irradiation and the influence of temperature, no hysteresis is observed when comparing curves 1 and 3 of the Figure 7. The absence of the hysteresis is an important feature of $Ch_{OH}$ structure of the $N^*_{tb}$-forming mixture, being an advantage for different optical applications.

*2.4 Frequency-dependence of selective Bragg reflection wavelength of the $Ch_{OH}$ structure*

In this subsection we will consider how the changes in the applied field frequency $f$ at certain value of voltage U affect the spectral features of $Ch_{OH}$ structure of the photosensitive $N_{tb}$-forming mixture. As it was shown in [39], the ChOH structure can be easily controlled by the frequency of the applied electric field, which can be a very promising approach for applications, in particular, as a way to create various keys for reliable encryption systems ensuring data security.

Figure 8a shows the tuning voltage dependence of the Bragg reflection wavelength $\lambda_{max}$ of the $Ch_{OH}$ structure which is in State 1 (*i.e.* 2$^{nd}$ order of Bragg diffraction [44]) for different frequencies $f$ of the applied field. Higher frequency $f$ of the applied field leads to the red shift of the Bragg reflection wavelength $\lambda_{max}$ as shown in Figure 8b.

Taking into account the frequency-dependent parameters in Equation 4 such as elastic constant $K_{22}(f)$ and $K_{33}(f)$, the average refractive index $\langle n \rangle = 1.7$ and the dielectric anisotropy $\Delta\varepsilon(f)$, which were analyzed in detail elsewhere [39], the calculated dependencies of $\lambda_{max}(U)$ for different values of frequency $f$ were obtained (Figure 8b).

We can conclude that the increase of frequency $f$ of the applied field should be accompanied by increasing of the tuning voltage $U$ applied to $Ch_{OH}$ structure to preserve the value of the Bragg reflection wavelength $\lambda_{max}$ (Figure 8c).

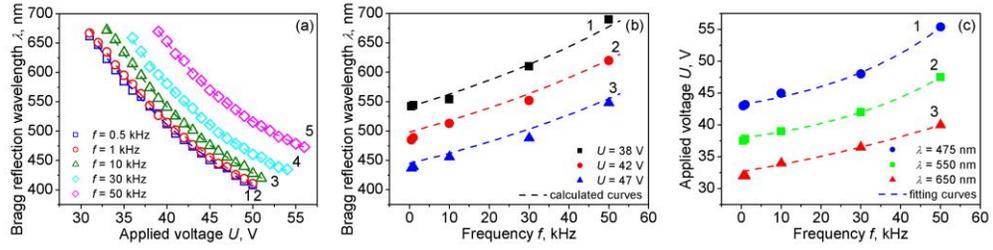

Fig. 8. Dependence of the Bragg reflection wavelength λ$_{max}$ of the LC cell filled by N*$_{tb}$-forming mixture on: (a) applied voltage $U$ for particular value of the frequency $f$: (1) - 0.5 kHz (opened blue squares), (2) - 1 kHz (opened red circles), (3) - 10 kHz (opened olive triangles) and (4) - 50 kHz (opened magenta diamonds) and (b) frequency $f$ of applied field at certain value of the applied voltage $U$: (1) – 37 V (solid black squares), (2) – 42 V (solid red circles) and (3) – 47 V (solid blue triangles). (c) Dependence of the tuning voltage $U$ applied to the LC cell on the frequency $f$ of the electrical field for particular Bragg reflection wavelength λ$_{max}$: (1) - 475 nm (solid blue circles), (2) - 550 nm (solid red squares) and (3) - 650 nm (solid red triangles).

The incidence angle dependence of selective BRL wavelength λ$_{max}$ of the Ch$_{OH}$ structure under electrical field (*i.e.* voltage $U$ and frequency $f$) is shown in Figure 9. At a certain value of the applied voltage $U$ and electric field frequency $f$, the increase of incidence angle $\theta$ leads to the red shift of reflected wavelength λ$_{max}$. Comparing Figure 9a, Figure 9b and Figure 9c we can conclude that with increasing of frequency $f$ at the certain applied voltage $U$ the decrease of deviation of the reflected wavelength λ$_{max}$ with increasing of incidence angle $\theta$ is observed. For instance, the changes in the BRL maximum wavelength observed with increased incidence angles are the smallest at frequency $f$ = 10 kHz (Figure 9c).

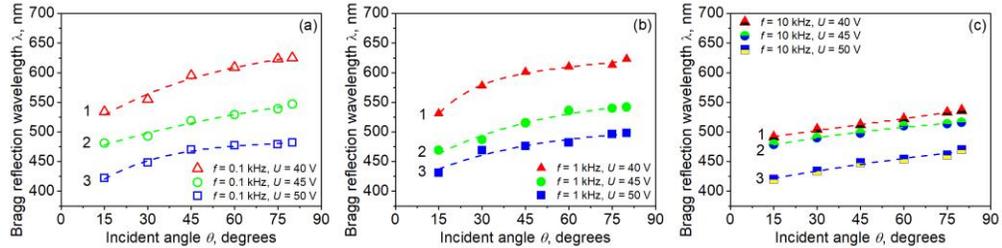

Fig. 9. Incidence angle dependence of the Bragg reflection wavelength λ$_{max}$ of the LC cell for the various values of applied voltage $U$ at frequency $f$: (a) 0.1 kHz; (b) 1 kHz and (c) 10 kHz.

3. **Conclusions**

In this work the effects of AC electric field of different voltage and frequency, as well as UV irradiation on the selective Bragg light reflection wavelength in the visible range of spectrum were studied for a photosensitive oblique helicoidal cholesteric structure containing the chiral azo-compound ChD-3816. It is shown that the use of photosensitive chiral compound as a component of the chiral twist-bend forming mixture leads to the possibility of light tuning of selective Bragg reflection wavelength of the $Ch_{OH}$ structure formed under electric field of voltage and frequency.

The photosensitive feature of the chiral twist-bend forming mixture can be used to recording of different information, which can be read out from $Ch_{OH}$ structure under specific conditions, *i.e.* at a certain value of voltage $U$, frequency $f$, time of irradiation $t_{irr}$ or their combination.

The major feature of $Ch_{OH}$ structure in comparison with $N^*$ phase is the absence of hysteresis under imposed changes of applied voltage, or in returning to the pre-irradiation state of the UV irradiated $N^*_{tb}$ mixture by means of Vis irradiation or thermal influence. This feature is important for the eventual use of the photosensitive $N^*_{tb}$ systems for different optical devices.